Introducing interdisciplinary science to second year undergraduates in a *Current Topics in Biophysics* course.

Authors: Stanislaw Jerzak (1) and Roger R. Lew (2)
((1) Department of Physics and Astronomy, York University, Canada, (2) Department of Biology, York University, Canada)

We offer second year students the opportunity to explore *Current Topics in Biophysics* in a course co-taught by a physicist and a biologist. The interdisciplinary course allows university students to engage in analytical thinking that integrates physics and biology. The students are either biophysics majors (50%) or from a diversity of science majors (about 30% life sciences). All will have taken first year courses in biology, physics and mathematics. The course is divided into: 1) The application of physical approaches to biological problems using case studies (how high can a tree grow? and biological pumps are two examples); 2) An introduction to physics concepts for which potential applications are explored (biophotonics and its application in fluorescence microscopy and photodynamic therapy is one example); and 3) Presentations from industry and university researchers who describe careers, research and clinical applications of biophysics. Over the six years the course has been offered, students have achieved a B average independent of declared major.  In course evaluations, students expressed a high level of satisfaction the course overall (91% were positive). However, students expressed reservations about the open-ended nature of assignments and test questions that approach biological problems using an analytical framework grounded in physics —the real challenge when implementing an open-ended problem solving approach to pedagogy.

# INTRODUCTION

The need for interdisciplinary approaches in scientific research exists at all levels of education, from primary and secondary schools (where science literacy is the primary goal [Project 2061, 1993]) to the training of biologists at the university level (National Research Council, 2003). Interdisciplinary approaches are exemplified by biophysics — integrating physics and the life sciences.

**Biophysics Curricula are Difficult to Define.** One of the major challenges in devising curricular content is the very nebulous definition of biophysics. Cell biologists may explore the nano-scale mechanics of molecular motors. Plant and animal physiologists tend to use a subset of physics (electricity and magnetism) and physical chemistry. Biochemists use physical tools, such as molecular spectroscopy. Ecologists may engage in modeling using ordinary differential equations and can use a variety of physical measuring techniques (telemetry for mapping animal movements, or ultrasonic recordings for studies of bat ecology are two examples that give some idea of the range). Structural biologists use magnetic resonance, x-ray crystallography and molecular spectroscopy to probe molecular structures of proteins and nucleic acids. Physicists on the other hand tend to study novel optical imaging techniques (Baker, 2011), quantum entanglement in photosynthesis (Sarovar et al., 2010), random walks (e.g., DNA entanglement and bacterial motility), and micro-hydrodynamics (Goldstein et al., 2007). Differences in scale also confound. For example, the physics used to explore the structure of a protein or ion transport across a membrane is quite different from the physics used to explore the motility of a bacteria (*versus* a fish or bird). The diversity of biophysics is not easily presented at a meaningful level in the early stages of an undergraduate program.

Creating an interdisciplinary biophysics curriculum is also challenged by the high level of specialization in modern research in both physics and the life sciences, which naturally leads to well-defined and narrow courses of study at the University level in the two disciplines. There have been serious attempts to reform first year curricula to create more interdisciplinary offerings. 'Physics for Life Sciences' textbooks are available (Appendix I) and there is a growing emphasis to ensure the authenticity of the biological problems that are used as examples and for assignments and tests (Watkins et al., 2012). Similarly, biology courses and programs have been reformed to emphasize quantitative aspects of Biology (Usher et al., 2010). A broader interdisciplinary scope has used integrated lecture and lab covering biology, chemistry and physics (Purvis-Roberts et al., 2009). At our university, biology and physics majors have the option of taking a first year course in Physics (including one course specifically for life scientists) or Biology as part of general science requirements. After first year, combined major programs are available (either a double major or major/minor combination), but only three students have graduated with a degree combining Biology and Physics in the past 14 years. Part of the reason may be a lack of interest in Physics on the part of university science students. There has been a long-term trend of declining enrollments in Physics, which has reversed since 1999 based on a survey of Physics departments in the United States (Nicholson and Mulvey, 2011). As a consequence of the decline, a number of Physics and Astronomy Departments in the

Province of Ontario initiated Biophysics (or Biological Physics) degree programs to enhance their undergraduate degree offerings. Such a program was also initiated at our university. Indeed, enrollments in Physics and Astronomy at our university have been increasing due to students enrolling in the biophysics major in their first year of university study (Figure 1).

In support of the new undergraduate degree program in biophysics, we developed a biophysics course at the second year level of university study. There were two objectives. The course serves as an introductory exploration of topics in biophysics for students in the biophysics degree program (who take biophysics courses in their third and fourth year). But in addition, the course was designed to encourage enrollment by students in other fields of study, especially life sciences. Thus, it was both a gateway to further studies in biophysics, and an opportunity to introduce students in other majors to interdisciplinary approaches in the sciences, providing them with an opportunity to integrate physics (and math) with the life sciences (Bialek and Botstein, 2004).

**Designing the *Current Topics in Biophysics* Course.** Fulfilling an interdisciplinary approach requires obligatory contributions from both physicists and biologists. Thus the course was co-taught by faculty members of the biology and the physics and astronomy departments to ensure comprehensive coverage from both a biological and physical perspective. To our knowledge, introductory biophysics courses are uncommon. Henderson and Henderson (1976) describe a team-taught introductory biophysics course at a similar level —second university year— that focused on the use of physics as a tool for understanding biological systems. The course covered the physics of sound, light, electricity and other topics in a biological context that complemented normal coverage in a first year physics course. We designed a format that introduces students to a range of current topics in biophysics, relying on a variety of sources (Appendix I). The underpinning theme was that biophysics is the application of physical approaches to biological problems. Guest lecturers (comprising about 1/8 of the lecture hours) came from both departments, or from industry and included both biologists and physicists who worked at the intersection of the two disciplines (for example, radiology and medical imaging).

A 'current topics' format allowed the freedom to devise a flexible curriculum that could be changed and modified without constraint (the current syllabus is shown in Appendix II). In keeping with that flexibility, emphasis was placed on introducing students to analytical thinking, in the form of open-ended assignments and tests that challenged the students to apply what they had learned to novel problems and concepts at the interface of biology and physics.

In this paper, we describe aspects of the curricular development and analytical approaches, and provide data on student outcomes.

## METHODS

**Quantitative Assessments of the Course.**

Data were analyzed in aggregate for the six years that the course has been offered. Course evaluations were given after one third of the course was complete and at its completion. In both instances the evaluation was administered during lecture time, and returned unsigned to ensure anonymity. The same evaluations were used during the six years the course has been taught. The evaluations used a Likert scale (of either 1–5 or 1–3) for specific questions focused on organization, pacing, difficulty, and interest. To evaluate outcomes, results for specific questions were used. For the first evaluation: 1) The course content was interesting (1–least positive to 5—most positive); 2) Rate this course overall (1–least positive, 5—most positive). For the second evaluation: 1) Course subject matter is 1—not very interesting, 2—moderately interesting, 3—highly interesting; Taken as a whole, would you recommend the course to another student with similar interests 1—Undecided, 2—No, 3—Yes. Values of 4 and 5 from the first evaluation were combined with values of 3 from the second evaluation to obtain the percentage of positive responses. The evaluations also provided for free-form anonymous comments by the students ("What did you like most/least about the course?" with spaces provided for "most" and "least"). The free form comments were collated by topic; the frequencies of positive or negative comments (as %, sample size) and representative comments are presented.

Academic outcomes in the course (in the form of final percentage scores) were compared by declared major and with the sessional grade point averages of the students at the completion of the academic year in which they took the course.

The data collection protocols and their use for research were approved by the Offices of Research Ethics (Human Participants Review Sub-Committee) and of Institutional Research and Analysis at York University.

## RESULTS

**Curricular Development—The Case Study Approach.**

In keeping with the general definition of 'physical approaches to biological problems', two formats were used. In one, a biological phenomenon was presented, and then the physical mechanisms at play were presented. This case study approach allowed foundational material to be presented as required, and encouraged analytical thinking by the students in the form of open-ended assignments and tests in which they were asked to apply the physical approaches to a new or different aspect of a biological problem. This type of approach works very effectively in a wide range of biological topics. Those that have been explored within the course include temperature regulation by organisms (requiring an understanding of the physics of radiative heat exchange and heat pumps (such as breathing and blood circulation in the case of mammals); bacterial motility

(flagellar nanomotors) (requiring an understanding of the microphysics that dominates in a low Reynolds number regime [Purcell, 1977 and Dusenberry, 2009]); biological pumps (highlighting the remarkably diverse array of physical mechanisms used to transport fluids) (requiring an understanding of hydrodynamics [Vogel, 2009]); and the height of a tree. The physics underlying the height of a tree provides a foundational introduction to mechanics, microfluidics, evaporative pumps and the tensile strength of water, and will be described in more detail.

The biological problem has its roots in evolution: The appearance of land plants (about 400 million years ago) that had to adapt to sparse availability of water on land, and in the context of competition for the light required to photosynthesize, the need to grow high (Niklas, 1997). The physical approach must be parsed (at the level of second year students) to biomechanics (mechanical support of the tree), micro-hydraulics (to transport the water to the top of the tree) and evaporative pumping (to pull the water to the top of the tree), and the constraints caused by the tensile strength of water (Figure 2). The students are therefore introduced to a diverse range of physics that would commonly be taught in separate courses.

The biomechanics is simplified to an introduction to the mathematical concept of the second moment of area and its application to an equation relating the mechanical strength of wood, the diameter of the tree column, and the height at which failure of the tree column will occur (McMahon, 1973).

The flow of water from the soil to the top of the tree is described in the context of the anatomy of the plants —emphasizing the xylem conduits for water flow and the stomates regulating water vapor release from the leaves. The physics behind this flow is defined by the hydraulic equation developed by Hagen and Poisieulle (Sutera and Skalak, 1993). Finally, the force required to 'pull' the columnar strand of water to the top of the tree is described in the context of the chemical potential of water transitioning from a liquid to vapor state. Thus, in one case study, students are introduced to the evolution of the biological form of a land plant that is explained by physics drawn from three distinct subjects. The actual physical limitation on tree height remains controversial. A recent hypothesis proposes that it is due to the breakage of the columnar strands of water due the limits caused by the tensile strength of water itself (Tyree, 2003).

At this point, the students are challenged to apply their knowledge to a related problem in the form of an assignment (and a term test), with the intent of presenting the students with an analytical challenge. For example, based on a speculative proposal by Freeman Dyson to bioengineer trees to grow on comets (Dyson, 2006), students were asked to determine how high a tree could grow on a comet given the marked difference in gravity. They could also have been asked to use the analytical tools of physics to test the popular misconception that osmosis could be the force causing water transport to the top of a 100 meter tree. In another example, students were asked to analyze how high a human could be on the basis of biomechanical constraints. A final example was to determine how large a mammalian heart pump would have to be to replace the evaporative pump of a tree. Students were expected to frame their answers in a realistic way; multiple physical approaches could be and were used. For example, in the case of a Dyson tree, they had to figure out how to estimate a comet's gravity, and its impact on structural strength and evapo-transpiration, amongst other things; for human height, they had to search for data on the strength of bones. Students were instructed that they "may wish to work together on the assignment, that is fine, but be sure that your assignment is in your own words". In fact, many of the students came for assistance with the assignments, mostly in groups of two or three students. They naturally engaged in peer-assisted learning by working in self-formed groups. For any of the assignments, the open-ended nature of the assignment meant that there was no clear and definitive answer. Rather, the assignments were graded on the basis of an appropriate choice of physical approaches, how realistic any necessary assumptions and/or estimates were, effort, and analytical insight. When asked soon after an assignment was returned whether "The grading system was fair", 82% (n=87) of the students selected 4 or 5 on a Likert scale of 1 (least positive) to 5 (most positive).

On the basis of student comments in course evaluations about assignments and tests (n=31), some students expressed a positive view of the open ended assignments ("the open-ended assignments were very interesting…" "…a nice change from most university assignments") (23%) but many (77%) found it "hard to answer open-ended questions" and difficult to know "what was expected on assignments [and tests]".

**Curricular Development —The Applications Approach.**

The second curricular approach presented foundational physics first, followed by its application to biological problems. The major topics were biophotonics, nuclear physics (and radiation biology) and magnetic resonance imaging. This approach went beyond classical physics to expose the students to quantum physics at a conceptual level. In all cases, the foundational physics was presented first, followed by examples of its application in biology. The major example was biophotonics. Concepts such as molecular orbitals, vibrational and rotational states, vibrational relaxation, singlet and triplet states, fluorescence and phosphorescence, were explained in the course, primarily qualitatively. The initial investment of time to introduce these quantum concepts paid off when students learned about biological utilization of light in photosynthesis and vision, and absorption properties of molecules functioning in oxygen transport and electron transfer processes. The diverse range of applications that were presented to students included fluorescence spectroscopy and microscopy, the molecular basis of vision and photodynamic therapies. Nuclear physics was also presented. Topics included: somatic and genetic damage of DNA due to radiation, deterministic and stochastic physiological effects of radiation, biological dose equivalent, radioisotopic labeling in biology and medicine and radionuclides in radiation tomography (PET—Positron Emission Tomography). One of the areas of biophysics that fascinates students is neurobiology; in particular, the unsolved puzzles surrounding the inner workings of the brain. The application of magnetic resonance to brain imaging provided an opportunity to teach students a hybrid of classical and quantum descriptions of the spin angular momentum and the associated magnetic dipole moment in magnetic field and how the spin relaxation times (longitudinal and transverse) can be used to obtain contrast for structural imaging of the brain. The difference in consumption of oxygen by active and inactive neurons is the basis of BOLD (Blood-Oxygen Level Dependent) f-MRI. A field trip to the f-MRI facility on campus provided a practical component for this portion of the course.

The assessments were a combination of work problems (assignments) and mostly conceptual questions in term tests and final exam. In assignments, students could be asked to calculate the energies of photons, the absorbance properties of biological pigments, heating effects of laser irradiation, etc. On the term test and final, the questions were either conceptual (What is a triplet state and how is it used in PDT? Why leaves are green? What are fluorophores and how are they used in fluorescence microscopy? Explain the role of the dichroic mirror in a fluorescence microscope?), or required calculations (Lambert-Beer law, effective half-lives in diagnostic medicine, biological dose equivalent, probability of getting cancer due to radiation exposure, spin-spin and spin-lattice relaxation times for different tissue).

**Curricular Development —Presentations by Biophysicists.**

An important innovation in this introductory course is the incorporation of four or five guest lectures delivered by experts in various areas of biophysics. Here, the topics vary dependent on availability of suitable presenters. From industrial and clinical settings, talks ranged from the challenges of diagnostic imaging in pathology to whole body imaging techniques in medical diagnostics (radiology, ultrasound, MRI, etc.). On the university research side, topics ranged from the use of particle beams in cancer therapy, electrical and functional MRI techniques in image processing, single cell analytical techniques and mass spectrometry. Student understanding of the material was evaluated by asking them to write summary gists (1–2 pages) of selected talks. The students were encouraged to use their own notes to write a brief summary of the "what, the how, and the why of the lecture", what excited them about the topic, and what they thought the impact would be. The gists served not only as a tool to test student understanding of the lecture topic, but also provided an opportunity to enhance their critical writing skills.

**Academic Assessment.**

The academic assessment was comprised of two term tests (worth 30%; for some years, the lower test score was weighted 10%, the higher test score was weighted 20%), assignments (two, worth 20%, one assignment was usually broken down into 2–4 mini-assignments), gists of selected guest lectures (two, worth 5% each) and a final exam (worth 40%). The average scores for each component (compiled over the 6 years the course has been taught) were: Assignments (84.6%), gists (77.6%), term tests (72.7%), and final exam (68.1%).

**Student Outcomes.**

By comparison with other 2[nd] year courses in either Biology or Physics, the enrollments in the *Current Topics in Biophysics* course were small, averaging 18 students (range: 10–24) in the six years it has been taught. Most 2[nd] year science courses in our institution have average enrollments of well over 100 students. Breakdowns by major and academic outcomes are shown in Table I. Academic performance in the course was unrelated to major.

The students rated the course highly based upon their responses to questions about how interesting the course material was, and their assessment of the course overall. The data were compiled from course evaluations presented after 12 lecture hours and at the end of the course. Overall, 87% of the students (n=141) found the course material interesting; 91% (n=139) rated the course highly overall. A common thread in the student comments was an appreciation for the current topics nature of the course (86%, n=51) ("Topics were very interesting and touched on many subject areas", the "material made me see many things from different perspective") and "having the chance to apply both physics and biology knowledge to practical situations". Some students expressed concerns that "the topics we cover are very random" and are "spread out and unrelated to one another" (14%).

Even with the "open-ended nature" of the course, student academic outcomes were good (a B average, defined as a good level of knowledge of concepts and/or techniques together with considerable skill in using them to satisfy the requirements of an assignment or course.). Their grades in the course were similar to their grades in other courses that they took during the year they took the *Current Topics in Biophysics* course. A regression analysis of their scores in *Current Topics in Biophysics* compared to their grade point averages for all courses taken during the same academic session revealed a strong correlation ($R^2 = 0.647$, n=99) (Figure 3). Thus, the academic assessment the *Current Topics in Biophysics* was in harmony with the academic standards in other courses the students took, independent of major.

**DISCUSSION**

The development of the *Current Topics in Biophysics* course occurred in the context of a shifting paradigm for science curriculum at our university: The genesis of a biophysics undergraduate degree program that integrates physics and the life sciences. The fact that overall enrollments in the Physics and Astronomy Department have increased as a consequence of the new degree program (with physics majors remaining unchanged) demonstrates a real interest on the part of first year University students to select interdisciplinary degree offerings when they are available. While the *Current Topics in Biophysics* prepares biophysics students for their upper-level biophysics courses, a secondary objective —to introduce students in other majors to interdisciplinary science— was also successful since 50% of the course enrollment has been non-majors, mostly students from the life sciences (30%). To address these two objectives, we developed a hybrid course that balanced case study approaches, applications approaches and presentations from biophysicists.

The term *Current Topics* often implies a survey of issues and controversies in any particular subject; sometimes it implies the latest research topics in science. We consider it to have a broader meaning: Biological problems that have not been resolved —some of these biological problems date back centuries— that benefit from an analytical approach using physics. The course academics are rigorous, and the students are challenged to exercise critical thinking in the form of open-ended assignments and tests in which they are asked to apply their knowledge to biological problems or situations they are not familiar with, and for which no definitive correct answer exists. After all, organisms have very effectively adapted to biological problems over thousands of millions of years, and they are seldom constrained by only one, simple physical limit. Hence, organisms have evolved open-ended solutions. Our approach is related to problem-based learning (Herreid, 1998). Students find the open-ended problems to be quite daunting, as evidenced by the concerns they expressed about the open-ended nature of the assignments and tests. Not knowing what to expect naturally results in anxiety and is very different from the normative pedagogical approaches of foundation courses commonly taught in the first and second year of university study. The assignments and grading scheme were designed to be a safety net for the students. Assignments and gists were graded more generously with significant weight placed on effort. A flexible weighting scheme was used for the term tests (the lowest mark was weighted less than the higher mark). We expect that this served to minimize grade anxiety on the part of the students, who did consider the grading to be fair (84%, n=87). Grade outcomes were similar to student grades in other courses. This reassures us that academic standards are harmonious, and suggests that strong quantitative/analytical reasoning goes hand-in-hand with better academic performance, independent of major.

Is this hybrid curricular structure an appropriate approach to introduce students to the intersection of physics and biology? We believe so. The biophysics majors are introduced to a wide-ranging variety of physical approaches that they will explore in subject-based courses in subsequent years. The non-majors are introduced to inter-disciplinary approaches and critical thinking in a small classroom environment that is quite different from the standard curricular approaches they will see in courses within their own majors at the $2^{nd}$ year level. A crucial requirement for those considering mounting a course like this is that the integration of physics and the life sciences is crucial, requiring an almost obligatory contribution by both physicists attuned to biology and life scientists attuned to physics.

An aspect of the course that we did not explore formally was the mathematical and physics underpinnings of the course. For students in biophysics and some of the other majors (physics and math), the mathematics should be simple and straightforward; for other majors, there is always a concern that students may feel 'math-challenged'. Only a few students (n=9) commented on the math aspect, some liked the "mathematical analysis", but others did not like the "amount of math and equations" and found it "a little too mathy". This did not create a barrier to success in the course, since academic outcomes were independent of major.

Offering a co-taught course is a realistic way to enhance the interdisciplinary experiences of students at the intersection of physics and biology. Based on our success, it seems clear that typical Life Sciences and Physics courses in the $1^{st}$ and $2^{d}$ year of university study could be co-taught in a more interdisciplinary way, a conclusion that could be easily extended to other science majors, such as chemistry.

**Accessing Materials.** An archived course website that includes assignments and test questions is available online at www.yorku.ca/planters/BPHS_2090_website/.

## ACKNOWLEDGEMENTS

We thank our biophysicist colleague Dr. Christopher Bergevin for his critical reading of the manuscript.

Table I. Summary of Enrollment and Academic Outcomes. Over the past four years, half of the students in the SC/BPHS 2090 course were majors in the biophysics undergraduate degree program (for whom the course is required, all were in their second year). Non-biophysics majors were in a variety of degree programs (and were in their $2^d$, $3^d$, or $4^{th}$ year of study). From 2008 to 2013, biology majors were about 22%, the other life sciences (kinesiology and biochemistry) accounted for 7%, physics and chemistry majors each accounted for about 5% of total enrollment from 2008 to 2013. 'Other' majors included psychology, science and technology, applied math, general science and undecided. Final scores in the course were compared between the biophysics majors and students in other majors; none of the differences were statistically significant using a 2-tailed t-test.

**Enrollment and Academic Outcome Summary (2008–2013)**

| Major | Total | % of Total | Final Score (mean ± SD) | t-test comparisons of Biophysics majors with other majors | |
|---|---|---|---|---|---|
| | | | | | *P–value* |
| Biophysics | 48 | 45.7 | 73.9 ± 11.7 | | |
| Biology | 23 | 21.9 | 78.8 ± 13.2 | | 0.142 |
| Kinesiology | 4 | 3.8 | 76.6 ± 1.4 | | 0.148 |
| Biochemistry | 3 | 2.9 | 85.5 ± 6.9 | | 0.082 |
| Chemistry | 5 | 4.8 | 64.5 ± 10.9 | | 0.126 |
| Physics | 6 | 5.7 | 64.5 ± 18.6 | | 0.069 |
| Other | 16 | 15.2 | 64.9 ± 15.4 | | 0.439 |
| Totals | 72 | | 73.3 | versus all other majors | 0.800 |

**FIGURES**

Figure 1. Enrollments in Physics and Astronomy: The impact of an interdisciplinary biophysics major program. First year students (freshman) entering the programs as physics majors (circles), biophysics majors (triangles) and the sum of both majors (squares) are shown over the past 13 years. Linear regressions are also shown. Since physics majors have remained relatively unchanged, the increased total entry enrollments are not a consequence of students selecting biophysics over physics, but rather the influx of new students into Physics and Astronomy from the first offering of the biophysics degree in 2007.

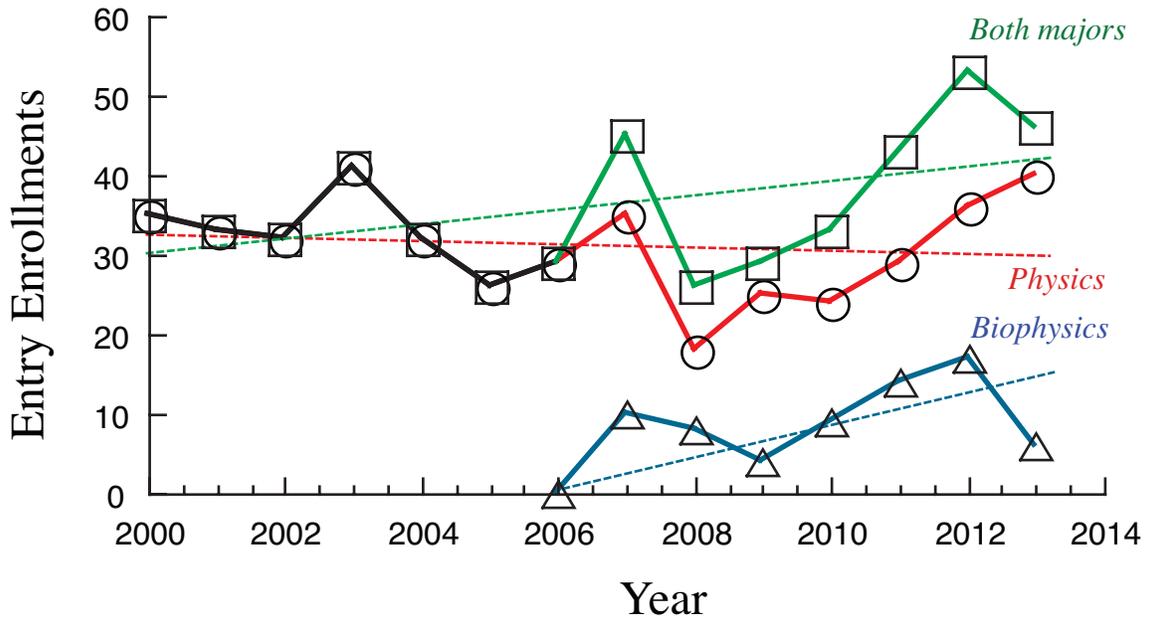

Figure 2. Schematic map of the curricular structure of the case study *How High can a Tree Grow*. The underlying theme is the presentation of physical approaches to biological problems. The physical approaches are drawn from 4 subjects in Physics (thermodynamics, fluid dynamics, mechanics and condensed matter). The equations describe the negative water potential ($\Psi_{wv}$) that pulls water to the top of a tree (a function of the relative humidity, %RH), the flow of water through xylem vessels ($J_v$) (a function of the pressure gradient ($\Delta p/l$), viscosity ($\eta$) and the xylem radius ($R$)), and the critical force at which a column fails ($F_{critical}$) which depends upon the strength of the wood ($E$), the effect of tree diameter ($I$) and the effective length of the tree ($L_{eff}$)) (McMahon, 1973). Examples of assessment questions are also shown.

# How High Can a Tree Grow?

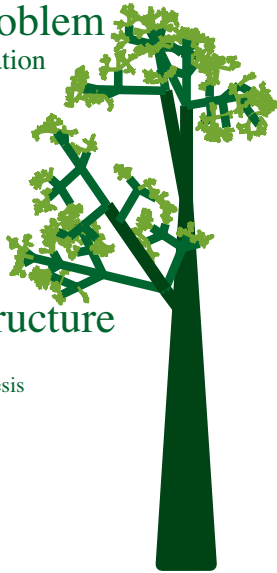

## Biological Problem
Evolution and adaptation
    Colonization of land
    Competition for light
Physical aspects
    Pumping water
    Water piping
    Structural support

## Biological Structure
Leaves
    Water for photosynthesis
Woody stem
    Xylem vessels
    Columnar structure
Woody roots
    Water uptake
    Structural foundation

## Physical Approach
Evaporative pump
(thermodynamics)

$$\Psi_{wv} = 135 \times \ln\frac{\%RH}{100}$$

Poiseuille flow
(fluid dynamics)

$$J_v = \left(\frac{\Delta p}{l}\right)\left(\frac{\pi}{8\eta}\right) R^4$$

Euler's column
(mechanics)

$$F_{critical} = \frac{EI\pi^2}{L_{eff}^2}$$

Tensile strength of water
(condensed matter)

## Assessment
Can osmotic pressure supply water to the top of a tree?

Does fluid mechanics constrain biological requirements in other biological processes (for example blood supply)?

What are the biomechanical constraints on heights of other organisms?

Explain experimental measurements of tensile strength in the scientific literature; tensile strength of fluids other than water

Figure 3. Academic outcomes: The *Current Topics in Biophysics* course *versus* sessional grade point average. The final scores of students in the *Current Topics in Biophysics* course are plotted *versus* their grade point average for all courses taken during the academic session in which they took *Current Topics in Biophysics*. For presentation, the data are binned by means ± SD (n=5) (to ensure anonymity), one outlier failing grade was removed. The sessional grade point averages were a good predictor of student performance in the course. The linear best fit to the un-binned data (n=99) was highly correlated ($R^2 = 0.647$). The grade equivalents are shown along with the percentage scores (y-axis) and the grade point average (x-axis).

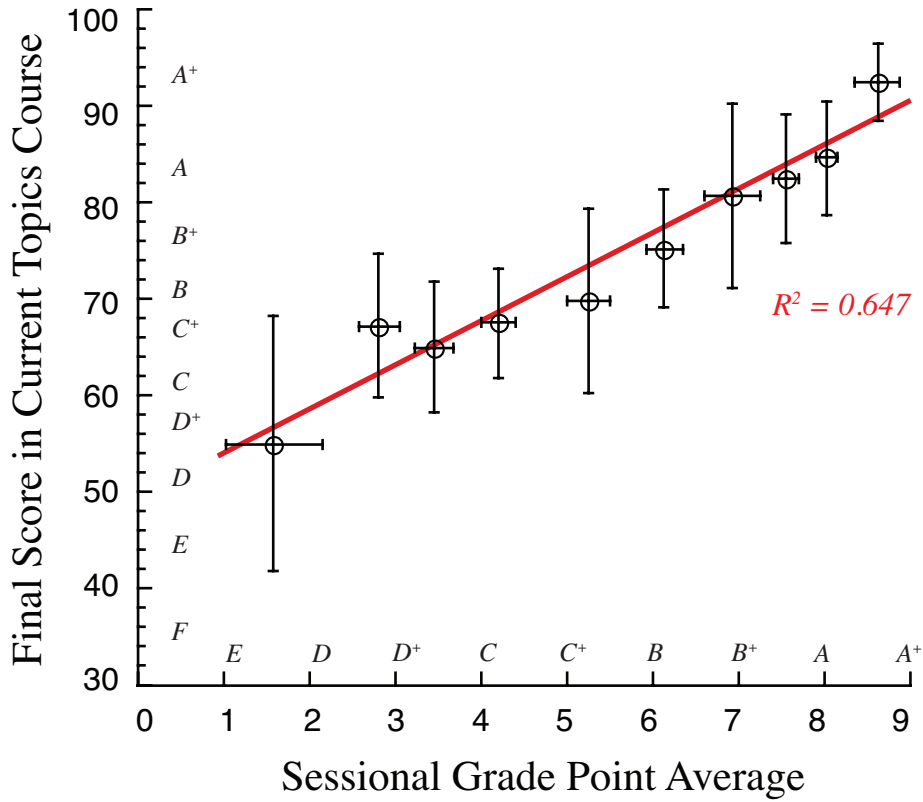